\documentclass[letterpaper, 10 pt, conference]{ieeeconf} 

\IEEEoverridecommandlockouts                              

\usepackage{float}
\usepackage{amsmath,amsfonts}
\usepackage{cite}
\usepackage{amssymb}
\usepackage{graphicx}
\usepackage{mathptmx}
\usepackage{times}
\usepackage{physics}
\usepackage{epstopdf}
\usepackage{mathtools}
\usepackage{subcaption}
\usepackage{comment}
\usepackage{xcolor}
\usepackage[export]{adjustbox}[2011/08/13]
\DeclareMathOperator*{\argmin}{arg\,min}
\DeclareMathOperator*{\argmax}{arg\,max}

\newtheorem{theorem}{\textbf{Theorem}}
\newtheorem{definition}{\textbf{Definition}}
\newtheorem{Remark}{\textbf{Remark}}
\newtheorem{Lemma}{\textbf{Lemma}}
\newtheorem{corollary}{\textbf{Corollary}}

\newtheorem{assump}{\textbf{Assumption}}

\title{\LARGE \bf
Finite time max-consensus for simultaneous target interception in switching graph topologies
}

\author{Kushal P. Singh$^{1}$, Aditya K. Rao$^{2}$ and Twinkle Tripathy$^{3}$
\thanks{*This work was partially funded by the project SRG/2022/001928.}
\thanks{$^{1}$Kushal P. Singh is a research scholar in the department of Electrical Engineering, Indian Institute of Technology Kanpur,
        Kanpur-208016, India
        {\tt\small kushalp20@iitk.ac.in}}%
\thanks{$^{2}$Aditya K. Rao is a research associate in the department of Electrical Engineering, Indian Institute of Technology Kanpur,
        Kanpur-208016, India
         {\tt\small adikrao@iitk.ac.in}}
\thanks{$^{3}$Twinkle Tripathy is an assistant professor with the Department of Electrical Engineering, Indian Institute of Technology Kanpur,
        Kanpur-208016, India
        {\tt\small ttripathy@iitk.ac.in}}%
}

\begin{document}

\maketitle
\thispagestyle{empty}
\pagestyle{empty}

\begin{abstract}
In this paper, we propose a distributed guidance law for the simultaneous interception of a stationary target. For a group of `n' heterogeneous pursuers, the proposed guidance law establishes the necessary conditions on static graphs that ensure simultaneous target interception, regardless of the initial conditions of the pursuers. Building on these results, we also establish the necessary conditions for achieving simultaneous interception in switching graph  topologies as well. The major highlight of the work is that the target interception occurs in finite time for both static and switching graph topologies. We demonstrate all of these results through numerical simulations.
\end{abstract}
\section{Introduction}
\label{Section:Introduction}
Recent advancements in defence systems and evasion strategies have made target interception using a single pursuer ineffective and prone to failure. To address this predicament, a recognised strategy is to intercept the target simultaneously with multiple pursuers. When a group of pursuers concur to intercept a target simultaneously, it overloads the target's defence system, making it vulnerable to the pursuers despite several countermeasures or defensive strategies deployed by the target.
The underlying guidance laws are generally designed in a distributed framework because of several advantages like scalability, efficiency in optimising and allocating resources, and avoidance of a single point of failure.

In the existing literature, the problem of simultaneous interception of a stationary target in static graphs is addressed in several ways. Guidance laws based on proportional navigation guidance (PNG) have proven to be very effective in target interception using a single agent. Similar cooperative guidance laws have been proposed for simultaneous target interception as well \cite{co_PPN,co_homing_guidance,co_two_stage_PN}. The authors in \cite{co_PPN} propose a two-stage guidance law for interception of a stationary target; the favourable initial conditions are generated first using a decentralised control law and, thereafter, pure proportional navigation guidance is used to govern the pursuers. In a similar framework, the authors in \cite{co_homing_guidance} propose a cooperative proportional navigation guidance law wherein the navigation gain of PNG is suitably varied for all the pursuers to achieve interception. By achieving consensus on the time-to-go estimates and  making these estimates realistic, the authors in \cite{co_two_stage_PN} propose two different guidance laws for simultaneous interception. The convergence analysis is shown using Lyapunov theory. Control theory-based methods have also been used for simultaneous target interception. The authors in \cite{co_multiple_sliding_surfaces} present a guidance law derived using multiple sliding surfaces; one sliding surface for every pair of neighbouring interceptors to synchronise their time-to-go and another global sliding surface to guide at least one interceptor to the target ensuring target capture. For better accuracy of the simultaneous target capture and faster convergence in impact times error, finite time consensus theory and supertwisting algorithm based sliding mode control algorithm is adopted in \cite{co_SMC_supertwisting_algorithm} to capture a stationary target.

Simultaneous target interception can also be accomplished by directing a team of pursuers to rendezvous at the target. In \cite{Pursuit_formations_of_unicycles}, the authors propose a feedback control strategy for achieving rendezvous with unicycles, specifying the necessary conditions for the control gains. The authors in \cite{Rendevous_Zheng} introduce a bearing-based control law that guides the pursuers to converge by reducing the perimeter of the polygon formed by their positions. A time-invariant discontinuous feedback control strategy is proposed in \cite{Rendevous_Demarogonas} for rendezvous in both positions and orientations. For a leader-follower approach, the authors in \cite{communication_delays} design a guidance law using feedback linearisation to achieve finite-time convergence of the followers' impact times with that of the leader. The authors in \cite{co_max_consensus,co_geometrical_rule} present a max- consensus based guidance law for simultaneous interception of a stationary target. A group guidance algorithm for target acquisition is proposed in \cite{co_unmanned_aerial_vehicles}, where an algorithm first drives the measured initial parameters to consensus and then a 3D proportional guidance algorithm independently guides the vehicles to the target.

The guidance laws discussed so far are designed for static graph topologies. However, in real-world scenarios, factors like device failures, the presence of obstacles and limited sensory ranges can often lead to switching graph topologies. In such graphs, consensus problems have been addressed using several approaches based on feedback linearisation \cite{dy_feedback_linearisation}, classical Nussbaum-type functions \cite{dy_delta_connected}, eventual consensus \cite{dy_eventual_consensus}, max consensus \cite{dy_max_consensus}, \textit{etc}. In \cite{dy_simultaneous_PNG}, a two-stage guidance law is introduced: initially, in the presence of a directed spanning tree, the law drives the pursuers to  consensus, after which each they are individually guided by PNG to intercept the target. Similarly, the authors of \cite{dy_simultaneous_with_angle} propose a guidance law comprising of two parts, the first part utilises terminal sliding mode control to achieve the desired impact angle, while the second part employs an integral sliding mode controller for simultaneous target capture. 
%
In this paper, our objective is to design a distributed guidance law for the simultaneous interception of a stationary target. Drawing inspiration from the works in \cite{co_geometrical_rule}, we propose a guidance law that allows a system of $n$ heterogeneous pursuers to achieve simultaneous interception of a stationary target. The guidance law is designed for pursuers modelled as a unicycle with constant speed. Under the proposed law, the trajectory of a pursuer is a concatenation of only straight-lines or circular arcs. Further, they are guided by a leader, which is simply the pursuer with the maximum estimated time of interception. By design, the leader node can change during the pursuit; nonetheless, it still drives the rest of the pursuers towards a simultaneous interception. The main contributions of this paper are as follows: 
\begin{enumerate}
    \item[i.] \textit{Finite time interception:} In the paper, we propose a max-consensus-based distributed cooperative guidance law to ensure simultaneous target interception in finite time regardless of their initial positions and heading angles. The challenging aspect here is to achieve consensus in the interception times of the pursuers in finite time which is ensured through the proposed law. 
    \item[ii.] \textit{Applicability to switching graph topologies:} Yet another advantage of the proposed law is that it is applicable to both static and switching graph topologies. Further, we show that consensus in the interception times can still be achieved even when leader is not globally reachable at all times.
    \item[iii.] \textit{Ease of implementation:} We propose the use of only straight line segments and circular arcs for the trajectories of the pursuers. Such a choice eases the design and implementability of the proposed law.
\end{enumerate}
The paper is structured as follows: Sec. \ref{Section:preliminaries} presents some required preliminaries. Sec. \ref{Section:Problem_Formulation} formulates the problem statement. Sec. \ref{section:cooperative_guidance_law} presents the proposed guidance law. Sec. \ref{sub:Acyclic} and Sec. \ref{sub:dynamic_graphs} illustrate the consensus in static and switching graphs, respectively. Sec. \ref{Section5} presents some simulation results and finally Sec. \ref{Section6} concludes the paper.
\section{Preliminaries}
\label{Section:preliminaries}
A static digraph is defined as $\mathcal{G}=(\mathcal{V},\mathcal{E})$, which is a directed graph consisting of a fixed finite number of nodes denoted by set $\mathcal{V}=\{1,2,...,n\}$ and a fixed finite number of edges having ordered pair of nodes denoted by set $\mathcal{E} \subseteq \mathcal{V}\text{x}\mathcal{V}$. 

The digraph $\mathcal{G}$ represents the communication topology between the pursuers in continuous time. The nodes represent pursuers, and the edges represent the communication between them. A directed edge $(i,j)\in \mathcal{E}$ denotes an outgoing edge from node $i$ to node $j$. The node $j$ is called an out-neighbour of node $i$ and node $i$ is called an in-neighbour of node $j$,  implying that the pursuer $i$ can only sense the pose of its out-neighbour $j$. Each node is assumed to contain a self-loop in the communication graph because each pursuer can sense its own pose.

The binary adjacency matrix $A=[a_{ij}]\in \{0,1\}^{n\times n}$ of digraph $\mathcal{G}$ is defined as $A_{ij}=1$ for $(i,j)\in \mathcal{E}$ and $a_{ij}=0$ otherwise.
A digraph $\mathcal{G}$ is called {\it{strongly connected}} if there exists a directed path from any node to any other node.
A digraph $\mathcal{G}$ is {\it{weakly connected}} if the undirected version of $\mathcal{G}$ is connected. 
A digraph $\mathcal{G}$ is {\it{complete}} if there exists a directed edge between every pair of nodes. 
A node $i$ is called {\it globally reachable} in a digraph if a directed path exists from every other node to node $i$.
$\mathcal{G}'=(\mathcal{V}',\mathcal{E}')$ is called {\it{subgraph}} of a digraph $\mathcal{G}=(\mathcal{V},\mathcal{E})$ if $\mathcal{V}' \subseteq \mathcal{V}$ and $\mathcal{E}' \subseteq \mathcal{E}$.
A subgraph $\mathcal{F}$ of $\mathcal{G}$ is called a {\it{strongly connected component}} of $\mathcal{G}$ if $\mathcal{F}$ is strongly connected and any other subgraph of $\mathcal{G}$ that contains $\mathcal{F}$ is not strongly connected.  

A switching digraph is defined as $\mathcal{G}(t)=(\mathcal{V}(t),\mathcal{E}(t))$, in which both the number of nodes and the number of edges may vary with time at discrete time intervals. The number of nodes can be changed by adding or removing nodes from the graph, and edges by changing both the number of nodes and the neighbours. The adjacency matrix changes instantly with the graph whenever switching occurs. Building upon these preliminaries, the following section formulates the problem statement.
%
\section{Problem statement}
\label{Section:Problem_Formulation}
In this work, we aim to design a distributed guidance law to achieve simultaneous interception of a stationary target $T$ using a group of heterogeneous autonomous pursuers.
\begin{figure}[ht]
    \centering
    \includegraphics[scale=0.75,center]{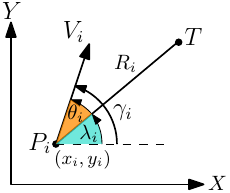}
    \caption{Basic engagement geometry}
    \label{fig:Fig1}
\end{figure}
The pursuers are modelled as unicycles whose equations of motion are given by eqn. \eqref{eq:kinematics}
\begin{subequations}
\label{eq:kinematics}
\begin{align}
\label{eq:(xidot)}
 \dot{x_i} & = V_i\cos{\gamma_i}  \\
 \label{eq:(yidot)}
 \dot{y_i} &= V_i\sin{\gamma_i} \\
 \label{eq:(gammaidot)}
 \dot{\gamma_i} & = a_i/V_i
 \end{align}
\end{subequations}
For every $i\in\{1,2,...,n\}$, $P_i$ denotes the position of pursuer $i$ as shown in Fig. \ref{fig:Fig1}, with coordinates $(x_i,y_i)\in \mathbb{R}^2$. Each pursuer $i$ has a constant linear speed $V_i$, controlled by lateral acceleration $a_i$, a heading angle $\gamma_i$ relative to a fixed reference frame, which is the angle between the heading direction of pursuer $i$ and the X-axis of a fixed reference frame, as well as a lead angle $\theta_i$, which is the angle between the heading direction of pursuer $i$ and the line-of-sight from $P_i$ to the target $T$, as illustrated in Fig. \ref{fig:Fig1}. The target $T$ has position coordinates $(x_t,y_t)\in \mathbb{R}^2$. Since the target remains stationary, each pursuer is assumed to be aware of the target's position for all time. 

While each pursuer can be commanded independently to reach the target $T$ at any feasible desired time, a cooperative framework can be designed to guarantee a simultaneous interception even when there are aberrations from the expected behaviour mid-course. This problem of simultaneous interception can also be addressed even if the communication channel between any two pursuers is temporarily unavailable during the engagement. 

\textit{Communication topology:} The digraph $\mathcal{G}(t)=(\mathcal{V}(t),\mathcal{E}(t))$ represents a time-varying communication, where nodes represent the pursuers and edges represent the directional communication between them. As defined in Sec. \ref{Section:preliminaries}, any entry $a_{ij}$ of the adjacency matrix could either be zero or one, which means any pursuer $i$ could either sense its out-neighbour or not. An example of this could be a camera-based sensor with a field of view. Depending on the field of view of the sensor, two different types of graphs can arise as discussed below:
\begin{itemize}
    \item \textit{Infinite field of view}: In this case, the out-neighbours of any pursuer cannot change due to an endless field of view. This always results in a static graph.
    \item \textit{Finite field of view}: The pursuers can move in and out of each other's field of view. This results in a switching graph which changes after finite intervals based on the locations of the pursuers. Owing to these finite intervals and finite field of view, the graph remains static for at least some finite time.
\end{itemize}
The underlying theory and the proposed guidance law for achieving simultaneous interception are discussed next.
\section{Cooperative guidance law}
\label{section:cooperative_guidance_law}
As cooperative guidance distributes its tasks across the network, cooperative guidance is better in terms of scalability, robustness, efficiency of operation, and resource allocation. Cooperative guidance also avoids the problem of a `single point of failure' and requires less sophisticated sensors than non-cooperative guidance. Therefore, in this paper, a group of pursuers coordinate their times of impact to intercept the target simultaneously. This covers the usage of circular and straight-line trajectories. The straight-line trajectories require zero control effort and the circular trajectories require constant lateral acceleration as input and hence are easy to implement, whose attributes are discussed next.
\label{Section3}
\subsection{Attributes of circular trajectory}
\begin{figure}[ht]
     \centering
     \begin{subfigure}[b]{0.45\linewidth}
     \centering
     \includegraphics[width=1.05\linewidth]{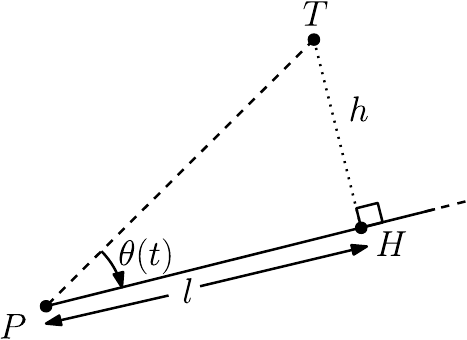}
     \caption{A straight line path}
     \label{fig:str motion}
     \end{subfigure}
     \hfill
     \begin{subfigure}[b]{0.42\linewidth}
     \centering    
     \includegraphics[width=0.82\linewidth]{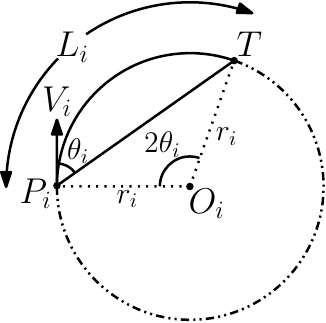}
     \caption{A circular path}
     \label{fig:Lemma2}
     \end{subfigure}
     \caption{Motion of the pursuer}
\end{figure}

Consider a pursuer $i$ initially at $P_i(x_i,y_i)$ with lead angle $\theta_i$. For this initial configuration, there is a unique circular path to $T$, requiring a lateral acceleration of
\begin{equation}
\label{eq:a_i_dash}
    {a_i}'=\frac{2{V_i}^2\sin{\theta_i}}{R_i}.
\end{equation}
The time required to follow this circular path is
\begin{equation}
\label{eq:t_tilde_i}
   \tilde{t}_i=\frac{R_i\theta_i}{V_i\sin{\theta_i}} \hspace{12pt} \text{where $\theta_i\in(-\pi,\pi)\backslash\{0\}$.} 
\end{equation}
This is referred to as \textit{estimated time of interception} in this paper. It represents the time needed for pursuer $i$ to reach the target on a circular trajectory from its current pose. To synchronise the impact times of all pursuers, in the absence of velocity regulation, those with shorter estimated times of interception must be delayed. The following Lemma shows that this delay can be achieved without any control effort if the pursuer maintains a straight-line path.
\begin{Lemma}
\label{Lemma:straight_line_Lemma}
Consider a set of $n$ pursuers with kinematics given by eqn. \eqref{eq:kinematics}. If a pursuer $i$, where $i\in \{1,2,...,n\}$, keeps on moving in a straight line, then $\tilde{t}_i(t)>0~\forall t\geqslant 0$ and it eventually increases monotonically.
\end{Lemma}
\begin{proof}
    Let the target $T$ be initially located at the origin and pursuer $i$ be at $(x_o,y_o) \in \mathbb{R}^2$, with a lead angle $\theta_o$, moving along the straight line $PH$ shown in Fig. \ref{fig:str motion}. From the geometry, the distance between $P$ and $T$(say $R_{PT}$) is given by,
    \begin{equation*}
        R_{PT}=\frac{h}{|\sin{\theta_t}|}.
    \end{equation*}
    where we know from Assumption \ref{assump:not_0_and_pi} that  $|\sin\theta_t|\neq 0$. Substituting $R_{PT}$ into eqn. \eqref{eq:t_tilde_i}, we obtain,
    \begin{equation*}
     \tilde{t}(t)=\frac{h\theta(t)}{v\sin\theta(t)|\sin\theta(t)|}.   
    \end{equation*}
    Differentiating $\tilde{t}_i$ w.r.t. time, we get, $    \Dot{\tilde{t}}_i(t)=1-2\theta(t)\cot{\theta(t)}.$ From Fig. \ref{fig:str motion}, we see that $\theta(t)\to\pm\pi$ for any pursuer moving on a straight line path. Therefore, $\Dot{\Tilde{t}}_i>0$ eventually and $\tilde{t}_i$ starts increasing monotonically.
\end{proof} 

While it is clear from Lemma \ref{Lemma:straight_line_Lemma} that moving in straight-line paths increases the estimated times of impact eventually, moving in circular paths might not have the same effect as elaborated next. 
\begin{Lemma}
\label{Lemma:-1_slope}
    Consider a pursuer $i$, with kinematics \eqref{eq:kinematics}, which is moving on a circular trajectory to intercept the target governed by the control input \eqref{eq:a_i_dash} where $i\in \{1,2,...,n\}$. Then, the rate of change of $\tilde{t}_i$ is given by $-1$.
\end{Lemma}
\begin{proof}
It follows from the geometry depicted in Fig. \ref{fig:Lemma2} that $L_i=2\theta_ir_i$. Since $r_i$ remains constant during the pursuer's circular motion, $$\frac{dL_i}{dt}=-2r_i\frac{d\theta_{i}}{dt}.$$ 
Now, since $(dL_i/dt)=V_i$, which is a constant, we can write $d\theta_i/dt=-V_i/(2r_i).$
Next, differentiating the relation $\tilde{t}_i=(L_i/V_i)$ w.r.t time gives $ d\tilde{t}_i/dt=(2r_i/V_i)(d\theta_i/dt).$ 
Substituting the expression for $(d\theta_i/dt)$, we get,
\begin{equation*}
 \frac{d\tilde{t}_i}{dt}=-1.
\end{equation*}
Hence, proved.
\end{proof}
%
\subsection{Guidance law}
Lemmas \ref{Lemma:straight_line_Lemma} and  \ref{Lemma:-1_slope} elaborate some valuable properties of straight line and circular trajectories in \textit{one-on-one} engagement scenarios. The following assumption is made to apply them for \textit{n-on-one} engagement scenarios.
\begin{assump}
\label{assump:not_0_and_pi}
 Throughout the paper, we assume that $\theta_i|_{t=0}=\theta_i(0)\notin \{0,\pi\}$ for all $i\in\{1,2,..,n\}$   
\end{assump}
Under Assumption \ref{assump:not_0_and_pi}, we propose the following guidance law, which ensures that the pursuers arrive at the target location at the same time,
\begin{subequations}
  \label{eq:control_law}
  \begin{align}
  \label{eq:control_law_part1}
   a_i &= 0 \hspace{18pt} \text{if $\tilde{t_i}<{\max}(\tilde{t}_{j}, j\in \mathcal{N}_{out}(i))$}\\
   \label{eq:control_law_part2}
   a_i &= a_i' \hspace{18pt} \text{if $\tilde{t_i}\geq {\max}(\tilde{t}_{i},\tilde{t}_{j}, j\in \mathcal{N}_{out}(i))$}
  \end{align}
\end{subequations}
where $\mathcal{N}_{out}(i)$ is the number of out-neighbours of pursuer $i$ in the communication graph $\mathcal{G}$ and $a_i$ is its lateral acceleration. In eqn. \eqref{eq:control_law}, the value of $\tilde{t}_j$ can be calculated using the pose information of pursuer $j$ which is available to pursuer $i$.
\begin{Remark}
\label{rem:1}
  In cases where Assumption \ref{assump:not_0_and_pi} does not hold at $t=0$, meaning $\theta_j\in \{0,\pi\}$ for $j\in \{j_1,j_2,...\}$. We apply a lateral acceleration $a_j=c$ for an arbitrarily short time $t_s$, where $c$ is some positive constant, to ensure $\theta_j \notin \{0,\pi \}$, while the remaining pursuers follow the guidance law given in eqn. \eqref{eq:control_law} for all times.

   By applying $a_j=c$, the affected pursuers follow a circular path, adjusting $\theta_j$ so that Assumption \ref{assump:not_0_and_pi} holds. Additionally, the time $t_s$ is chosen to be arbitrarily small enough that the order of estimated times of interception remains unchanged, meaning if $\tilde{t}_1<\tilde{t}_2<...<\tilde{t}_n$ at $t=0$, the same order holds at $t=t_s$ i.e. $\tilde{t}_1(t_s)<\tilde{t}_2(t_s)<...<\tilde{t}_n(t_s)$.
\end{Remark}
The guidance law guarantees that if Assumption \ref{assump:not_0_and_pi} holds for all $i$, $\theta_i$ is not zero throughout the engagement. The same has been highlighted in the next Lemma.
\begin{figure*}[ht]
     \centering
     \begin{subfigure}[b]{0.3\textwidth}
         \centering
         \includegraphics[width=\textwidth]{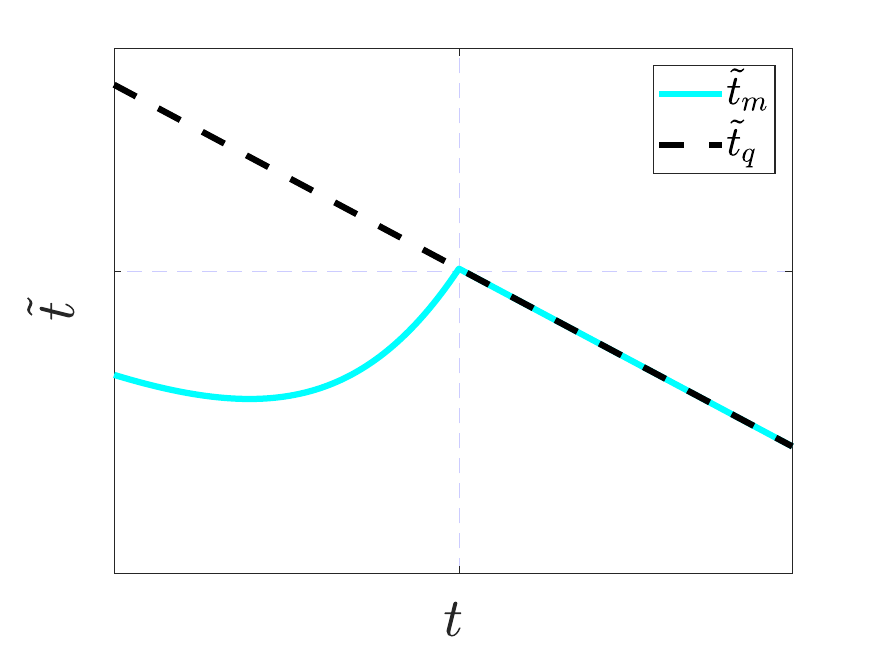}
         \caption{$q$ - circle, $m$ - straight line}
         \label{fig:i-1<i}
     \end{subfigure}
     \hfill
     \begin{subfigure}[b]{0.3\textwidth}
         \centering
         \includegraphics[width=\textwidth]{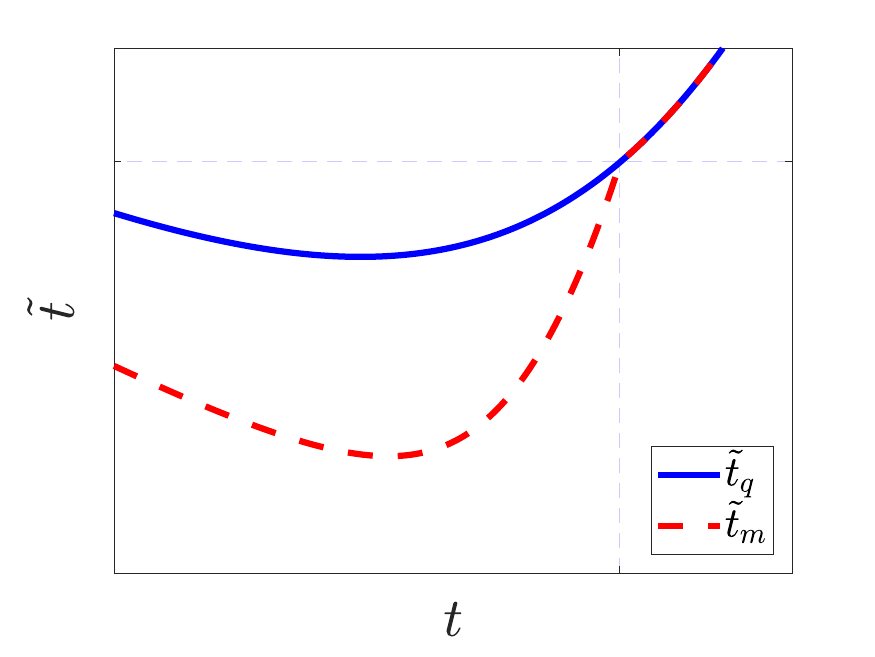}
         \caption{$q$ and $m$ - straight lines}
         \label{fig:ss}
     \end{subfigure}
     \hfill
     \begin{subfigure}[b]{0.3\textwidth}
         \centering
         \includegraphics[width=\textwidth]{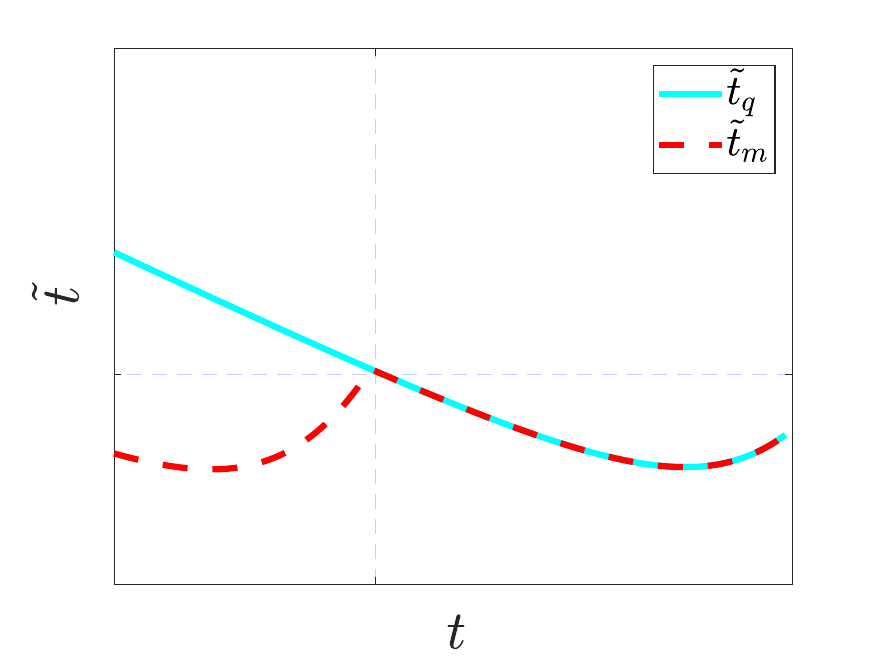}
         \caption{$q$ and $m$ - straight lines}
         \label{fig:leema4}
     \end{subfigure}
        \caption{The figures show the plots of $t\sim\tilde{t}$ for different motions of pursuers $q$ and $m$.}
\end{figure*}
\begin{Lemma}
\label{Lemma:theta_i_nonzero}
If Assumption \ref{assump:not_0_and_pi} holds for any pursuer $i$, then $\theta_i$ does not go to zero throughout the engagement.
\end{Lemma}
\begin{proof}
Eqn. \eqref{eq:a_i_dash} gives the lateral acceleration required to move on a circular path. From eqn. \eqref{eq:a_i_dash}, we can write $\sin{\theta_i}={a_i}'{R_i}/2{V_i}^2$. In this expression, $V_i$ and ${a_i}'$ have non-zero values. So, the only possibility for $\sin{\theta_i}$ to become zero is at interception when $R_i$ becomes zero. Also, a pursuer moving on a straight line path will always have a non-zero value of $\theta_i$ if Assumption \ref{assump:not_0_and_pi} holds. 

The resulting trajectory under \eqref{eq:control_law} is composed of circular arcs and straight lines. For any realisation of the trajectory, $\theta_i$ does not become zero in any subsection (circle or straight line) of the trajectory. Hence, proved.
\end{proof}

\textit{Idea behind the proposed guidance law:} Lemmas \ref{Lemma:straight_line_Lemma} and \ref{Lemma:-1_slope} show that the target can be intercepted on a circular trajectory with a constant input. Furthermore, we can delay the pursuers with smaller estimated times of impact by moving them in a straight line with no control effort. The guidance law is shown in eqn. \eqref{eq:control_law} ensures that if any pursuer $i$ has a shorter estimated time of impact than any of its out-neighbours, then pursuer $i$ should take a straight line path, eventually increasing its estimated time of impact $\tilde{t}_i$ until it equals to the estimated time of impact of out-neighbour having the largest estimated time of impact.

Based on the guidance law, we now discuss a few scenarios regarding when and how consensus and simultaneous interception can be achieved in a multi-agent environment. One of many such interactions is explained in the next Lemma.
\begin{Lemma}
\label{Lemma:larger_remains_larger}
 Consider the scenario wherein the $m^{th}$ pursuer, $m\in \mathcal{V}$, follows all of its out-neighbours and $q$ is the out-neighbour with the largest initial estimated time of impact, i.e. $ \tilde{t}_q(t)=\max_{p\in \mathcal{N}_{out}(m)}(\tilde{t}_p(t))$. If $\tilde{t}_q(0)\geqslant \tilde{t}_m(0)$, then
  the following relation holds,
  \begin{align}
  \label{eq:q_geq_m}
      \tilde{t}_q(t)\geqslant \tilde{t}_m(t)~ \forall ~t\geqslant 0.
  \end{align}
\end{Lemma}
\begin{proof}
  Under the assumption \ref{assump:not_0_and_pi}, the pursuer $q$ at $t=0$, depending upon its out-neighbours in the communication graph, could either be moving on a circle or a straight line. But, the pursuer $m$ always starts in a straight line path as $\tilde{t}_q(0)\geqslant \tilde{t}_m(0)$. Consequently, $\tilde{t}_m(t)$ eventually increases as shown in Lemma \ref{Lemma:straight_line_Lemma}. The following cases can then arise:
  \begin{itemize}
      \item[-] \textit{The pursuer $q$ moves in a circular path}: If the pursuer $q$ moves in a circular path at $t=0$,  $\tilde{t}_q$  decreases as shown in Lemma \ref{Lemma:-1_slope}. Since, $\tilde{t}_m$ is increasing, after a finite time $\tau_1$, $\tilde{t}_m=\tilde{t}_q$. Then, pursuer $m$ satisfies eqn. \eqref{eq:control_law_part2} and both the pursuers start moving in circular paths. This is depicted in Fig. \ref{fig:i-1<i}.
      \item[-] \textit{The pursuer $q$ moves on a straight line such that $d\tilde{t}_q/dt>0$}: $\tilde{t}_q$ increases at $t=0$ as its derivative w.r.t. time is positive and $\tilde{t}_m$ increases as per Lemma \ref{Lemma:straight_line_Lemma}. If the rate of increase of $\tilde{t}_m$ is less than or equal to $\tilde{t}_q$, then $\tilde{t}_m$ always remains less than $\tilde{t}_q$ because of initial conditions. And if the rate of increase of $\tilde{t}_m$ is greater than that of $\tilde{t}_q$, then as soon as $\tilde{t}_m$ equals $\tilde{t}_q$, pursuer $m$ satisfies eqn. \eqref{eq:control_law_part2}, it results in a decrease of $\tilde{t}_{m}$ at a rate of $-1$ as stated in Lemma \ref{Lemma:-1_slope}. Pursuer $m$ moves momentarily on a circle, and $\tilde{t}_{m}$ becomes less than $\tilde{t}_q$, so the pursuer $m$ now again moves in a straight line path and then again in a circular path and so on. So, $\tilde{t}_m$ never moves away from $\tilde{t}_q$ once they become equal, as shown in Fig. \ref{fig:ss}.
      \item[-] \textit{The pursuer $q$ moves on a straight line such that $d\tilde{t}_q/dt<0$}: Since $d\tilde{t}_q/dt<0$, $\tilde{t}_q$ starts decreasing at $t=0$ and we have already established that $\tilde{t}_m$ is increasing, so there comes a time when they both become equal as shown in Fig. \ref{fig:leema4}. Thereafter, $\tilde{t}_m$ never moves away from $\tilde{t}_q$ as explained in the previous case.
  \end{itemize}
Note that the path of pursuer $q$ is a concatenation of sub-paths that always falls under either one of the three cases. For each one, $\tilde{t}_q(t)\geqslant \tilde{t}_m(t)$ holds. Hence, it holds for all time.
%
\end{proof} 
\begin{Remark}
   It should be noted that pursuer $q$, in this case, is not fixed and may change over time, but regardless of whether or not $q$ remains fixed, eqn. \eqref{eq:q_geq_m} always holds.  
\end{Remark}
Based on Lemma \ref{Lemma:larger_remains_larger}, the following definition entitles the pursuer with the largest value of estimated time of impact for all time.
\begin{definition}[Leader]
\label{def:leader}
We define a pursuer $l$ as the leader of the group of $n$ pursuers if the following relation holds,
\begin{equation*}
    \tilde{t}_l=\underset{1\leqslant i\leqslant n}{\max} ~\tilde{t}_i.
\end{equation*} 
\end{definition}
Note that the communication strategies in this paper are leaderless, meaning that none of the pursuers is a designated leader initially. However, based on the initial conditions, one (or more) of the pursuers acts as a leader(s). The remaining pursuers synchronise their impact times with the leader for simultaneous interception. The following Lemma shows how the pursuer with a larger impact time than its out-neighbours synchronises its impact time with them.
\begin{Lemma}
\label{Lemma:finite_time_convergence}
    Consider a pursuer $m\in\mathcal{V}$ and its out-neighbour $q$, which has the largest estimated time of impact among its out-neighbours at any given moment
     i.e., 
     $ \tilde{t}_q(t)=\max_{p\in \mathcal{N}_{out}(m)}(\tilde{t}_p(t))$.
  %
    If pursuer $q$ moves in a straight line and $\tilde{t}_q(0)<\tilde{t}_m(0)$, then there exists a finite time $\tau\geqslant 0$ such that $\tilde{t}_q(\tau)=\tilde{t}_m(\tau)$.
\end{Lemma}
\begin{proof}
%
%
As pursuer $q$ moves in a straight line at $t=0$, $\tilde{t}_q$ eventually increases according to Lemma \ref{Lemma:straight_line_Lemma}. And since pursuer $m$ satisfies eqn. \eqref{eq:control_law_part2} initially, it moves in a circular path. Then, $\tilde{t}_m$ reduces with a slope of $-1$, as stated in Lemma \ref{Lemma:-1_slope}. Because $\tilde{t}_m$ decreases with time and $\tilde{t}_q$ eventually increases with time, there exists a finite time $t=\tau$ when they become equal. Hence, proved.
 \end{proof}
 
 Lemma \ref{Lemma:larger_remains_larger} and \ref{Lemma:finite_time_convergence} highlight the local behaviour of pursuers in a large network under the proposed guidance law given in eqn. \ref{eq:control_law}. With these local properties, the following sections discuss the consensus results for pursuers having static and switching communication topologies.
\section{Consensus in static graphs}
\label{sub:Acyclic}
In this section, we explore the problem of simultaneous interception of a stationary target. The pursuers are located such that the communication graph remains static. Let $\mathcal{G}=(\mathcal{V},\mathcal{E})$ be a static digraph with set of nodes $\mathcal{V}=\{1,2,...,n\}$ and set of edges $\mathcal{E} \subseteq \mathcal{V}\text{x}\mathcal{V}$.

The existence of a globally reachable node in a graph plays a pivotal role in achieving consensus amongst pursuers for both static and switching graphs. These globally reachable nodes may drive the kinematics of pursuers towards a consensus. Owing to this, we require the leader (Definition \ref{def:leader}) to be {\it{globally reachable}}. The following Lemma helps us establish consensus in static digraphs with the above assumption.
\begin{Lemma}
\label{Lemma:less than leader}
For every pursuer $j$ other than the leader $l$, 
\begin{align}
\tilde{t}_j(t)\leq\tilde{t}_l(t)    
\end{align}
for all $t>0$ where $j\in\{1,2,...n\},~j\neq l$.
\end{Lemma}
\begin{proof}
Consider some pursuer $j\in\{1,2,...n\},~j\neq l$. Suppose pursuer $k_1$ is the pursuer with the largest $\tilde{t}$ among the out-neighbours of $j$ such that $\tilde{t}_j\leq\tilde{t}_{k_{1}}$. Similarly pursuer $k_2$ is the pursuer with the largest $\tilde{t}$ among the out-neighbours of $k_1$ such that $\tilde{t}_{k_1}\leq\tilde{t}_{k_{2}}$, this goes on till we find a pursuer $k_m$ such that $\tilde{t}_{k_m}$ is strictly greater than all of its out-neighbours. Using Lemma \ref{Lemma:larger_remains_larger}, we get the relation:
$\tilde{t}_j\leq\tilde{t}_{k_1}\leq\tilde{t}_{k_2}...\leq\tilde{t}_{k_m}~~\forall ~t>0.$

  Now we choose the pursuer with the largest $\tilde{t}$ among the out-neighbours of $k_m$ and denote it by $k_{m+1}$. Clearly, $\tilde{t}_{k_m}>\tilde{t}_{k_{m+1}}$. Proceeding similarly from $k_{m+1}$, we find a series of largest $\tilde{t}$ out-neighbours which terminate at the leader $l$ as it is globally reachable and has the largest $\tilde{t}$. We get  $\tilde{t}_{k_{m+1}}\leq\tilde{t}_{k_{m+2}}...\leq\tilde{t}_l~~\forall~t>0$. Consequently, $\tilde{t}_{j}\leq\tilde{t}_{k_1}...\leq\tilde{t}_{k_{m}}>\tilde{t}_{k_{m+1}}..\leq\tilde{t}_{l}$. In other words, $\tilde{t}_j\leq\tilde{t}_{k_{m}}$ and $\tilde{t}_{k_{m+1}}\leq\tilde{t}_l$ for all $t>0$.
  
  Using Lemma \ref{Lemma:finite_time_convergence}, there exists a finite time $\tau$ such that  $\tilde{t}_{k_m}$ monotonically decrease for all $t<\tau$ at the same rate as $\tilde{t}_l$ and $\tilde{t}_{k_m}(\tau)=\tilde{t}_{k_{m+1}}(\tau)$. Thus, 
\begin{equation}
    \tilde{t}_{k_m}~~
    \begin{cases}
        \leq~~\tilde{t}_{l}, & \text{if } t<\tau\\
        \leq~~\tilde{t}_{k_{m+1}}\leq\tilde{t}_{l}, & \text{if } t\geq\tau
    \end{cases}
    \label{eq:inequality km}
\end{equation}
Since $\tilde{t}_j(t)\leq\tilde{t}_{k_{m}}(t)$, using above relation, $\tilde{t}_j\leq\tilde{t}_{l}$ holds true for all $t>0$. 

Note that more than one node like $k_m$ may exist, such that $\tilde{t}_{k_m}$ is greater than that of its out-neighbours. We can write a relation similar to eqn. \eqref{eq:inequality km} for all such nodes with possibly different values of $\tau$. Hence, proved.
\end{proof} 

Lemma \ref{Lemma:less than leader} shows that none of the pursuers can have a higher $\tilde{t}$ than the leader for all time. This information can be used to calculate the final impact time of the leader at any time $t$.
\begin{Lemma}
    The leader $l$ of the group moves on a circular trajectory throughout the target engagement and its estimated time of interception is given by,
    \begin{equation}
        \tilde{t}_l(t)=\tilde{t}_l(0)-t.
        \label{eq:evolution of leader}
    \end{equation}
    \label{lem:evolution of leader}
\end{Lemma}
\begin{proof}
    We know from Lemma \ref{Lemma:less than leader} that $\tilde{t}_l$ remains the largest for all time $t\geqslant 0$. So eqn. \eqref{eq:control_law_part2} always holds for the leader $l$, and it moves in a circular trajectory throughout the engagement. Thus from Lemma \ref{Lemma:-1_slope}, $\tilde{t}_l$ decreases monotonically at a constant rate of $-1$. The expression for the estimated time of impact follows directly. Hence, proved.
\end{proof}

It is clear from Lemma \ref{lem:evolution of leader} that the leader intercepts the target at $t=\tilde{t}_l(0)$ and is unaffected by the behaviour of its neighbouring pursuers. The following results show that none of the other pursuers can reach the target before $t=\tilde{t}_l(0)$. 
\begin{Lemma}
\label{Lemma:cannot_before_leader}
For every pursuer $j$ other than the leader $l$, 
\begin{align}
\tilde{t}_j(t)>0 
\end{align}
for all $t\in[0,\tilde{t}_l(0))$ where $j\in\{1,2,...n\},~j\neq l$.
\end{Lemma}
\begin{proof}
A pursuer $j$ moves either on a straight line or a circle under the guidance law \ref{eq:control_law}. Given Assumption \ref{assump:not_0_and_pi}, if the path of $j$ is a straight line for any duration throughout the engagement, $\tilde{t}_j(t)$ is greater than zero (Lemma \ref{Lemma:straight_line_Lemma}). 

We now prove by contradiction that pursuer $j$ cannot intercept the target before the leader $l$ by moving on a circle. Let us suppose it intercepts the target $T$ at some time $\tau_j<\tilde{t}_l(0)$ which implies $\tilde{t}_j(\tau_j)=0$. In the current setup, we know from Lemma \ref{Lemma:-1_slope} that $\tilde{t}_j$ can be reduced to zero only by moving on a circular path. So, the pursuer $j$ must have been on a circular path for some duration just before the interception.

Consider an arbitrary time interval $[\tau_j-\epsilon_j,\tau_j]$ with $0<\epsilon_j<\tau_j$ where pursuer $j$ moves on a circular path {and ${\tilde{t}_j(t)<\tilde{t}_l(t)}$}. Additionally, its estimated time of impact is greater than or equal to all of its out-neighbours i.e., $\tilde{t}_k(t)\leqslant \tilde{t}_j(t)$ for all $t\in[\tau_j-\epsilon_j,\tau_j]$ where $k\in\mathcal{N}_{out}(j)$. Since pursuer $j$ intercepts the target at $t=\tau_j$, as per inequality $\tilde{t}_k(t)\leqslant \tilde{t}_j(t)$ obtained from eqn. \eqref{eq:control_law_part2}, the $\tilde{t}_k$ should also be zero at $t=\tau_j$ for all $k$. { Clearly from Lemma \ref{Lemma:straight_line_Lemma}, this is not possible if pursuer $k$ is moving on a straight line path.} Thus, the only possible conclusion is that all the out-neighbours of pursuer $j$ also move on a circular path for some duration just before interception and $\tilde{t}_k(t)$ monotonically decreases for $t\in[\tau_k-\epsilon_k,\tau_k]$ and becomes zero at $t=\tau_k\leqslant\tau_j$. 

Similarly, we can argue that any out-neighbour $k_1$ of any pursuer $k\in\mathcal{N}_{out}(j)$ also moves on a circle with $\tilde{t}_{k1}(t)\leqslant\tilde{t}_{k}(t)$ in the time interval $[\tau_{k}-\epsilon_{k},\tau_{k}]$ where $\epsilon_{k}$ is chosen to be arbitrarily small. Since $l$ is globally reachable, we can proceed similarly to find a pursuer $r$ that moves on a circle to intercept the target at $t=\tau_r\leqslant\tau_j<\tilde{t}_l(0)$ and has $l$ as its out-neighbour with $\tilde{t}_{l}(t)\leqslant\tilde{t}_{r}(t)$ {for ${t\in[\tau_r-\epsilon_r,\tau_r].}$} This is a contradiction since $\tilde{t}_r<\tilde{t}_l$ (Lemma \ref{Lemma:less than leader}) for all $t\leqslant \tau_r<\tilde{t}_l(0)$. Thus, pursuer $j$ cannot move on a circular path and intercept the target T at time $t=\tau<\tilde{t}_l(0)$. Thus, $t_j>0$ for all $0<t<\tilde{t}_l(0)$ irrespective of its motion throughout the engagement. Hence, proved.
\end{proof}

It is interesting to note that if $\tau_i=\tilde{t}_l(0)$ for all $i\in\mathcal{V}\setminus\{l\}$, the contradiction would not have arisen and all the pursuers would have moved on a circular path \textit{simultaneously} along with the leader and intercepted the target at $t=\tilde{t}_l(0)$. We now present conditions for achieving consensus among the pursuers.
\begin{theorem}
\label{Theorem_consensus_static}  
 Consider a system of $n$ pursuers, with kinematics \eqref{eq:kinematics}, in pursuit of a stationary target $T$. Let pursuer l be the leader of this group as defined in \ref{def:leader}. Under the guidance law \eqref{eq:control_law}, all the pursuers are guaranteed to intercept $T$ simultaneously at time $t_f=\tilde{t}_l(0)$.
\end{theorem}
\begin{proof}
It follows from Lemma \ref{Lemma:cannot_before_leader} that  $\tilde{t}_j(t)>0$  $\forall t\in[0,\tilde{t}_l(0))$ and from Lemma \ref{Lemma:less than leader} that $\tilde{t}_j(t)\leqslant \tilde{t}_l(t)$ $\forall t\geqslant 0$. Thus, 
\begin{equation}
    0<\tilde{t}_j(t)\leqslant \tilde{t}_l(t)
    \label{eq:bound on t_tilde}
\end{equation}
for all $t\in[0,\tilde{t}_l(0))$ and $j\in \mathcal{V}$. We know from Lemma \ref{lem:evolution of leader} that the right-hand side bound of inequality \eqref{eq:bound on t_tilde} goes to zero at $t=\tilde{t}_l(0)$ for all $j\in\mathcal{V}$. Thus, all $\tilde{t}_j$ s goes to zero at $t=\tilde{t}_l(0)$ achieving simultaneous interception. Hence, proved. 
\end{proof} 

We conclude from Theorem \ref{Theorem_consensus_static} that all the pursuers synchronise their estimated times of impact with that of the leader. Hence, max-consensus is achieved in $\tilde{t}\text{s}$. Further, $\tilde{t}$ is constrained to evolve in a bounded region. Using eqn. \eqref{eq:bound on t_tilde}, it follows that, for every $i\in\mathcal{V}$, $\tilde{t}_i(t)$ is constrained to evolve in the region $\Omega:=\{(t,\Tilde{t}_i(t))\in\mathbb{R}^{2}_{+}~|~\tilde{t}_i(t)+t-\Tilde{t}_l(0)\leqslant 0\}$. 
%
%
%
    \begin{figure}[ht]
        \centering
        \includegraphics[scale=0.6]{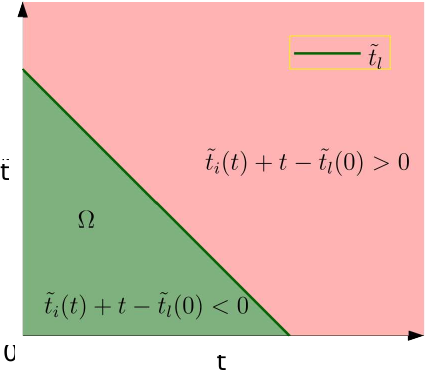}
        \caption{Bounds on evolution of $\tilde{t}$}
        \label{fig:forbidden}
    \end{figure}
\begin{corollary}
\label{corollary:cir_bef_int}
If a group of $n$ pursuers is on its way to intercept the target as per eqn. \eqref{eq:control_law}, then there exists a finite time $t_\alpha<\tilde{t}_l(0)$, such that all pursuers take circular trajectories for all time $t\in [t_\alpha,\tilde{t}_l(0)]$.
\end{corollary}
\begin{proof}
We have shown in Lemma \ref{Lemma:cannot_before_leader} that none of the pursuers can intercept the target $T$ before $t=\tilde{t}_l(0)$ while moving on a circular path. Thus, it must switch to a straight line path with $\tilde{t}$ eventually increasing. From Lemma \ref{Lemma:less than leader}, when $\tilde{t}$ becomes equal to $\tilde{t}_l$, both of the pursuers proceed on circular paths to achieve simultaneous interception. This behaviour is exhibited by all pursuers other than the leader. Then, there must exist $t_\alpha<\tilde{t}_l(0)$ at which the last pursuer achieves consensus with the leader. For all $t>t_\alpha$, the other pursuers are already moving on circular paths. Hence, proved.
\end{proof}

In an arbitrary graph with the leader as the globally reachable node, as discussed in Corollary \ref{corollary:cir_bef_int}, the switches between straight lines and circular arcs occur at least once for every pursuer other than the leader. Hence, for a complete graph, we can precisely determine the number of switches by applying Theorem \ref{Theorem_consensus_static} and Corollary \ref{corollary:cir_bef_int} as given below.
\begin{corollary}
      Consider a set of $n$ pursuers connected over a complete graph $\mathcal{G}$. Then, every pursuer other than the leader(s) switches its trajectory exactly once from a straight line to a circular arc.
\end{corollary}
We have established the conditions required to achieve consensus in $\tilde t$s, leading to simultaneous target interception in static graphs. Next, we explore how simultaneous interception can still be achieved when the graphs change with time.
\section{Consensus in switching graphs}
\label{sub:dynamic_graphs}
In practical scenarios, factors such as limited sensory ranges, device failures and the presence of obstacles can adversely hinder the communication between the pursuers, which leads to switching graphs. In this section, we investigate the conditions required for simultaneous interception in such graphs. 

A globally reachable leader's existence is necessary for a static graph to achieve consensus in $\tilde{t}$s, as discussed in Sec. \ref{sub:Acyclic}. Hence, we have the following result for switching graphs.
\begin{theorem}
\label{Theorem:consensus_dynamic_connected}
    Consider a switching graph of $n$ pursuers such that the leader remains globally reachable at all times. Then, the pursuers are guaranteed to intercept the target simultaneously under the guidance law defined in eqn. \eqref{eq:control_law}.
\end{theorem}
\begin{proof}
     In the given framework, as discussed in Sec. \ref{Section:Problem_Formulation}, the switching graph $\mathcal{G}(t)$ always varies from one static graph to another. Then, the variation of $\mathcal{G}(t)$ can be represented as a sequence $\{\mathcal{G}_z,~z\in \{1,2,...\}\}$, wherein every $\mathcal{G}_z$ is a static graph and the switch to $\mathcal{G}_z$ occurs at $t=t_z$. Consequently, we get a sequence of switching time instances $\{t_z\}$. All the results of Sec. \ref{sub:Acyclic} hold in $[t_z,t_{z+1})$ as $\mathcal{G}_z$ contains a globally reachable node. Therefore,  $\forall t\in[t_z,t_{z+1})$, 
     $0<\tilde{t}_i (t)\leqslant\tilde{t}_l(t)$,
     for all $i\in \mathcal{V}$ and $z\in \mathbb{Z}^+$ as per Theorem \ref{Theorem_consensus_static}. Since the right-hand side bound of above inequality goes to zero at $t =  \tilde{t}_l(0)$, the system keeps progressing towards the consensus and finally attains simultaneous interception. Hence, proved.
\end{proof}

\begin{figure*}[ht]
    \centering
    \includegraphics[scale=1,center]{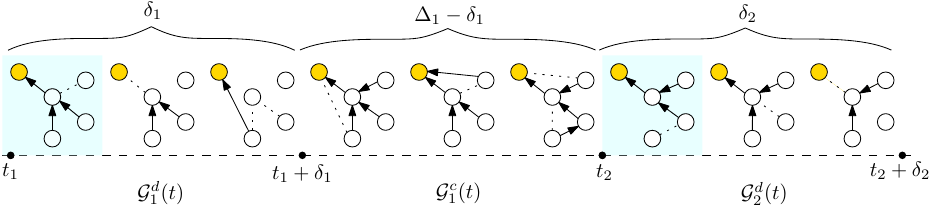}
    \caption{The figure illustrates the graphs for the time interval $[t_1,t_2+\delta_2)$ for a system of $5$ pursuers. The yellow node denotes the leader and the cyan graph represents the first graph where the switch occurs from $\mathcal{G}^c_{i-1}(t)$ to $\mathcal{G}^d_i(t)$.  Dotted edges indicate edges removed from the previous graph. For $t\in[t_1,t_1+\delta_1)$, all graphs are of the form $\mathcal{G}^d_1(t)$, where the leader is not globally reachable. For $t\in[t_1+\delta_1)$, the graphs are of the form $\mathcal{G}^1_c(t)$ where the leader remains globally reachable at all times.}
    \label{fig:gc_gd_figures}
\end{figure*}

In a general scenario, the leader may not always be globally reachable at all times in a switching graph. To address such scenarios, we define a sequence of time instances $\{t_m\}$ at which the leader $l$ loses its property of global reachability. We further define the following graphs in the time interval $[t_m,t_{m+1})$ where $m\in\mathbb{Z}^+$:
\begin{itemize}
    \item for a time interval $[t_m,t_m+\delta_m)$, the leader $l$ is not globally reachable and we denote the communication graph as $\mathcal{G}^d_m(t)$, and
    \item for a time interval $[t_m+\delta_m,t_{m+1})$, the leader $l$ is globally reachable and we denote the communication graph as $\mathcal{G}^c_m(t)$, 
\end{itemize}
where $\delta_m>0$ is the duration for which $\mathcal{G}^d_m(t)$ exists. Further, we denote $\Delta_m=t_{m+1}-t_m$. In Fig. \ref{fig:gc_gd_figures}, we show a sample sequence of graphs $\{\mathcal{G}^d_1(t),\mathcal{G}^c_1(t),\mathcal{G}^d_2(t),...\}$ existing for the sequence of duration $\{\delta_1,\Delta_1-\delta_1,\delta_2,...\}$.
%
%
While $\mathcal{G}^d_m(t)$ (or $\mathcal{G}^c_m(t)$) is inherently time-varying with changing edges and nodes, the leader $l$ is not (or is) globally reachable at any instance for the duration of $\delta_m$ (or $\Delta_m-\delta_m$). Note that the graphs may vary be due to the addition or subtraction of nodes or edges. We first analyse the switching graphs with fixed nodes and changing edges, and the necessary conditions to achieve consensus in $\tilde{t}$s in such scenarios. 
\subsection{Switching graphs with fixed nodes}
\label{subsub:only_edge_changes}
Let there be a sequence of digraphs $\mathcal{G}^c_m(t)=(\mathcal{V},\mathcal{E}(t))$ and $\mathcal{G}^d_m(t)=(\mathcal{V},\mathcal{E}(t))$ with fixed nodes. The following theorem presents the necessary conditions for such switching graphs to achieve consensus in $\tilde{t}_i$s .
\begin{theorem}
 \label{theorem:consensus_dynamic}
    Consider a system of $n$ pursuers under the guidance law  \eqref{eq:control_law}. Let $t_m$ denote the time instance at which the leader loses its global reachability for the $m^{th}$ time and the communication graph changes from $\mathcal{G}^c_{m-1} (t)$ to $\mathcal{G}^d_m(t)$. If 
    \begin{equation}
    \label{eq:dyn_graph_discontinuity}
    \delta_m < \min_{i\in\mathcal{V}}~\tilde{t}_i(t_m),
    \end{equation}
    for every $m$, then the system is guaranteed to achieve consensus in $\tilde{t}_i$ s.
\end{theorem}
\begin{proof}
    In the duration $\Delta_m-\delta_m$, the communication graph is $\mathcal{G}^c_m(t)$. As elaborated in Theorem \ref{Theorem:consensus_dynamic_connected}, the pursuers then progress towards a consensus in $\tilde{t}_i$s.
    
    In the duration $\delta_m$, the communication graph is $\mathcal{G}^d_m(t)$. Let $k=\underset{i}{\text{argmin}}\{\tilde{t}_i(t_m)\}$. If pursuer $k$ has an out-neighbour at $t=t_m$, $\tilde{t}_k$ moves on a straight line and does not intercept the target before the leader, as seen in Lemma \ref{Lemma:cannot_before_leader}. Otherwise, $\tilde{t}_k$ decreases with slope $-1$ as per the guidance law defined in \eqref{eq:control_law} and would require the time of $\tilde{t}_k(t_m)$ to intercept the target. But pursuer $k$ cannot intercept the target if $\delta_i<\tilde{t}_k(t_m)$. All other pursuers require more time than pursuer $k$ to intercept the target. Thus, eqn. \eqref{eq:dyn_graph_discontinuity} ensures that none of the pursuers intercepts the target before the leader even when the leader is not globally reachable, i.e., $\tilde{t}_i>0$ for all $t\in[t_m,t_m+\delta_m]$ and $i\in\mathcal{V}$.
    
    Again at $t=t_m$, we know from Theorem \ref{Theorem:consensus_dynamic_connected} that any pursuer $k\in \mathcal{V} \backslash \{l\}$ has $\tilde{t}_k(t_m)\leq\tilde{t}_l(t_m)$. If a path from node $k$ to node $l$ exists in $\mathcal{G}^d_m(t)$ during $\delta_m$, then $\tilde{t}_k(t)\leq\tilde{t}_l(t)$ for all $t\in[t_m,t_m+\delta_m]$ as per Lemma \ref{Lemma:less than leader}. If a path from node $k$ to node $l$ does not exist, then either the node $k$ has an out-neighbour $y$ such that $\tilde{t_k}(t_m)<\tilde{t_y}(t_m)$ or  $\tilde{t_k}(t_m)\geqslant \tilde{t_y}(t_m)$ or node $k$ is a singleton. If $\tilde{t_k}(t_m)<\tilde{t_y}(t_m)$, then $\tilde{t_k}   \leqslant \tilde{t}_y$ in the duration $[t_m,t_m+\delta_m)$ as per Lemma \ref{Lemma:larger_remains_larger}  and $\tilde{t}_y\leq\tilde{t}_l$. Otherwise $\tilde{t}_k$ decreases monotonically with a slope of $-1$, thus ensuring that $\tilde{t}_k$ always remains less than $\tilde{t}_l$ for all $t\in [t_m,t_m+\delta_m)$.
    
    Eqn. \eqref{eq:dyn_graph_discontinuity} ensures that none of the pursuers $k$ intercept the target before the leader and $\tilde{t}_k\leqslant \tilde{t}_l$ for the duration of $\delta_m$ for all $m$. Therefore, if $\delta$ is designed using eqn. \eqref{eq:dyn_graph_discontinuity}, the system attains a consensus in $\tilde{t}$s and the pursuers achieves simultaneous interception. Hence, proved.
\end{proof}
\begin{Remark}
\label{remark: fastest decrease of tilde t}
    An important observation about the evolution of $\tilde{t}_i$ in the given framework is that if a pursuer moves on a circle, the rate of decrease of $\tilde{t}_i$ (which is -1) is the fastest. Any alternate trajectory (any permutation of the circle and straight line paths) results in a slower decrease of $\tilde{t}_i$.
\end{Remark}
The choice of $\delta_k$ in Theorem \ref{theorem:consensus_dynamic} is crucial, as it ensures that no pursuer intercepts the target too early when the leader is not globally reachable. This is because the pursuer with the smallest $\tilde{t}$ may follow a circular path to target if it is a sink node in $\mathcal{G}^d_k(t)$. But when this pursuer has an out-neighbour in $\mathcal{G}^d_k(t)$, it moves on a straight line path, dismissing the possibility of target interception in the $\delta_k$ duration. Under such a scenario, it might be possible to impose a larger bound on $\delta_k$. The next corollary presents an interesting result on the first pursuer to intercept the target when the graph does not have a globally reachable node. 
%
\begin{corollary}
\label{corr:fastest to zero}
    Consider a static communication graph of a system of pursuers represented by $\mathcal{G}=(\mathcal{V},\mathcal{E})$ and $C\mathcal{(G)}=(\mathcal{V}_C,\mathcal{E}_C)$ be its condensation graph. For all $k\in\mathcal{V}_C$, let $\mathcal{T}(k)$ be the set of all nodes in $\mathcal{G}$ condensed to $k$. Let $\mathcal{S}\subseteq\mathcal{V}_C$ be the set of singleton and sink nodes in $C(\mathcal{G})$. Then, the pursuer $f=\underset{i\in P}\argmin~\tilde{t}_i(0)\in\mathcal{V}$ where
    \begin{equation}
    P=\underset{i\in\mathcal{S}}\bigcup~\left\{\underset{j\in\mathcal{T}(i)}\argmax~\tilde{t}_j(0) \right\}
    \end{equation}
    is the first pursuer to intercept target $T$ under the guidance law \eqref{eq:control_law} at time $t=\tilde{t}_f(0)$.
\end{corollary}
\begin{proof}
    Let $i\in\mathcal{S}\subseteq \mathcal{V}_c$ be some sink/singleton node in $C(\mathcal{G})$ that represents a strongly connected component in $\mathcal{G}$. Then, the corresponding subgraph of $\mathcal{G}$ does not have any out-neighbors and, hence, is not affected by any other nodes in $\mathcal{G}$ under guidance law \eqref{eq:control_law}. From Theorem \ref{Theorem_consensus_static}, all the pursuers synchronise their $\tilde{t}$s with the local leader $s=\underset{j\in\mathcal{T}(i)}\argmax~\tilde{t}_j(0)$. We construct the set $P$ using all such local leaders $s\in\mathcal{V}$. 
    For all $s\in P$, there are no out-neighbors, and every $\tilde{t}_s(t)$ decreases monotonically at slope $-1$ throughout the engagement. Let $f=\underset{i\in P}\argmin~\tilde{t}_i(0)$. Pursuer $f$ intercepts the target $T$ first among all pursuers in $P$.

    Now consider some pursuer $p\notin P$. We show that any such pursuer cannot intercept the target before the pursuer $f$. Let $\tilde{t}_p(0)\geq\tilde{t}_f(0)$. Now, the fastest way for any $\tilde{t}_p(t)$ to decrease is at a rate of $-1$ when the pursuer moves continuously on a circular path (Remark \ref{remark: fastest decrease of tilde t}). Thus, it can be shown that $\tilde{t}_p(t)\geq\tilde{t}_f(t)$ throughout the engagement, and $p$ cannot intercept the target $T$ before $f$. Let $\tilde{t}_p(0)<\tilde{t}_f(0)$. Every $p\notin P$ has a path to one or more nodes in $P$. Let $s\in P$ serve as its local leader. From Lemma \ref{Lemma:cannot_before_leader}, $\tilde{t}_p$ converges to $\tilde{t}_s$ and intercepts the target at $t=\tilde{t}_s(0)$. Since $\tilde{t}_f(0)\leqslant\tilde{t}_s(0)$, all such $p$ will intercept the target with or after $f$. Hence, proved.
\end{proof}

It is clear from Corollary \ref{corr:fastest to zero} that the sinks of $C(\mathcal{G}(t))$ control the evolution of the system of pursuers. By following a line of proof similar to that of Theorem \ref{theorem:consensus_dynamic}, we can arrive at a larger bound on $\delta_k$ as explained in the next result.
%
\begin{theorem}
\label{Theorem:consensus_const_sink}
    Consider a system of $n$ pursuers under the guidance law  \eqref{eq:control_law}. Let $t_m$ denote the time instance when the leader loses its global reachability for the $m^{th}$ time and the communication graph changes from $\mathcal{G}^c_{m-1} (t)$ to $\mathcal{G}^d_m(t)$. If $\mathcal{G}^d_m(t)$ varies such that its sinks/singletons cannot increase in number during $\delta_m$, the system achieves consensus for
    \begin{equation}
        \delta_m<\tilde{t}_f(t_m)
    \end{equation}
    where pursuer $f$ is defined in Corollary \ref{corr:fastest to zero}.
\end{theorem}
%
The primary difference between Theorem \ref{theorem:consensus_dynamic} and \ref{Theorem:consensus_const_sink} is the choice of $\delta_k$. While the former states the condition on $\delta_k$ using $\tilde{t}$s of all the nodes, the latter uses only the sink and singleton nodes. This can result in a larger choice of $\delta_k$, allowing the  graphs to remain disconnected for a possibly longer duration. The caveat being that new sink or singleton nodes cannot get formed in any $\delta_k$ interval. Next, we discuss the case of addition or deletion of nodes from the network.
%
\subsection{Switching graphs with changing nodes}
\label{subsub:only_node_changes}
Consider a scenarios in which the new nodes can be added or removed from the network. Owing to this, the leader might not remain globally reachable, or the leader may even change. So, the following corollary specifies the conditions that must hold to ensure consensus in $\tilde{t} \hspace{0.02cm}s$.
\begin{corollary}
\label{cor:addition/removal}
   Consider a system of $n$ pursuers under the guidance law  \eqref{eq:control_law} with a switching graph $\mathcal{G}(t)$. Simultaneous interception can be ensured under \eqref{eq:control_law} if the following conditions hold. 
   \begin{itemize}
    \item If a node is removed from $\mathcal{G}(t)$, then the leader of the resulting graph must be globally reachable.
    \item If a node $b$ is added to $\mathcal{G}_c(t)$, then,
    \begin{itemize}
        \item[-] $b$ must have at least one out-neighbor in $\mathcal{G}^c(t)$ if it is not the leader in the resulting graph,
        \item[-] else, $b$ should be a globally reachable node.
    \end{itemize}
    \item If a node $c$ is added to $\mathcal{G}^d_m(t)$ at time $t_a$ where $t_m\leqslant t_a \leqslant \delta_m+t_m$, then
    $t_a+\tilde{t}_c(t_a) \geqslant t_m+\delta_m$ must hold where $t_m$ and $\delta_m$ are as defined in Theorem \ref{theorem:consensus_dynamic}.
\end{itemize}
\end{corollary} 
If the aforementioned conditions are met along with those mentioned in Theorem \ref{theorem:consensus_dynamic}, then the system achieves consensus. The proof of Corollary \ref{cor:addition/removal} is along the same lines as that of Theorem \ref{theorem:consensus_dynamic}. After guaranteeing the consensus in static 
\begin{figure*}[ht]
     \centering
     \begin{subfigure}[b]{0.23\textwidth}
        \centering
        \includegraphics[width=\linewidth,center]{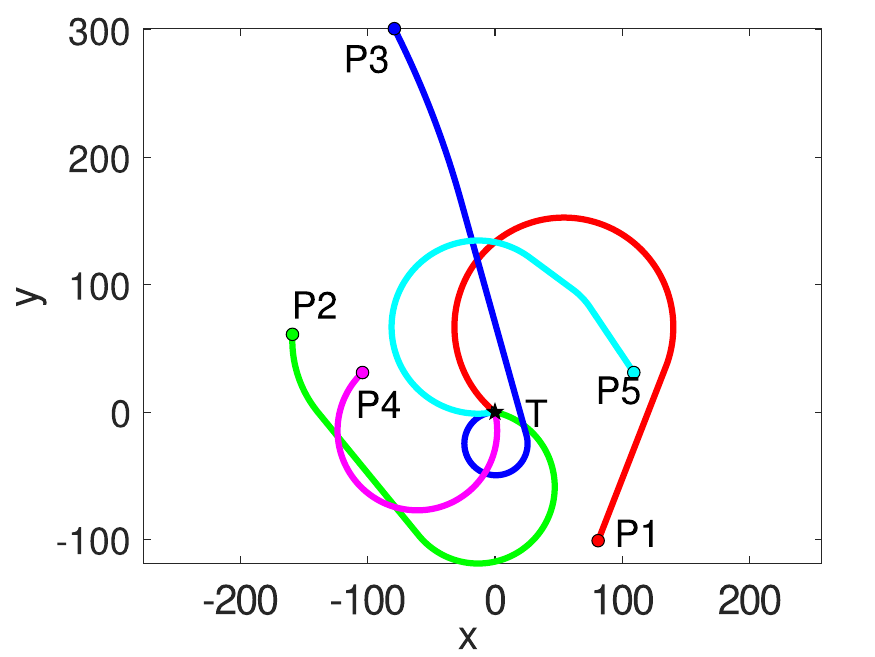} 
        \caption{Case $1$}
        \label{fig:Case1}
     \end{subfigure}
     \hfill
     \begin{subfigure}[b]{0.23\textwidth}
        \centering    
        \includegraphics[width=\linewidth,center]{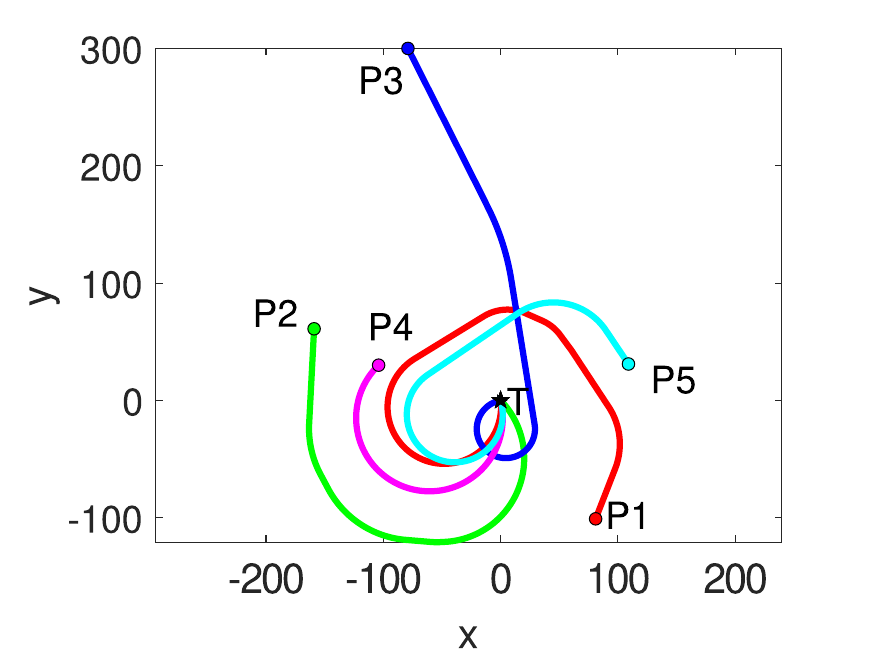}
        \caption{Case $2$ (without sinks)}
        \label{fig:trajectory_case_2}
     \end{subfigure}
     \hfill
     \begin{subfigure}[b]{0.23\textwidth}
    \centering 
    \includegraphics[width=\linewidth,center]{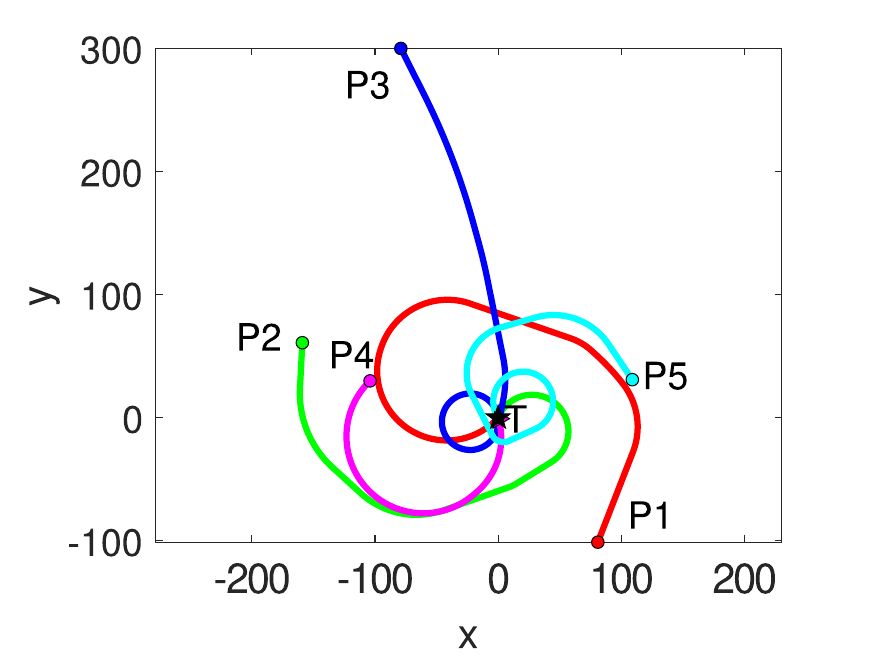}
    \caption{Case $2$ (with sinks)}
    \label{fig:Trajectories_Case3}
     \end{subfigure}
     \hfill
    \begin{subfigure}[b]{0.23\textwidth}
        \centering 
        \includegraphics[width=\linewidth,center]{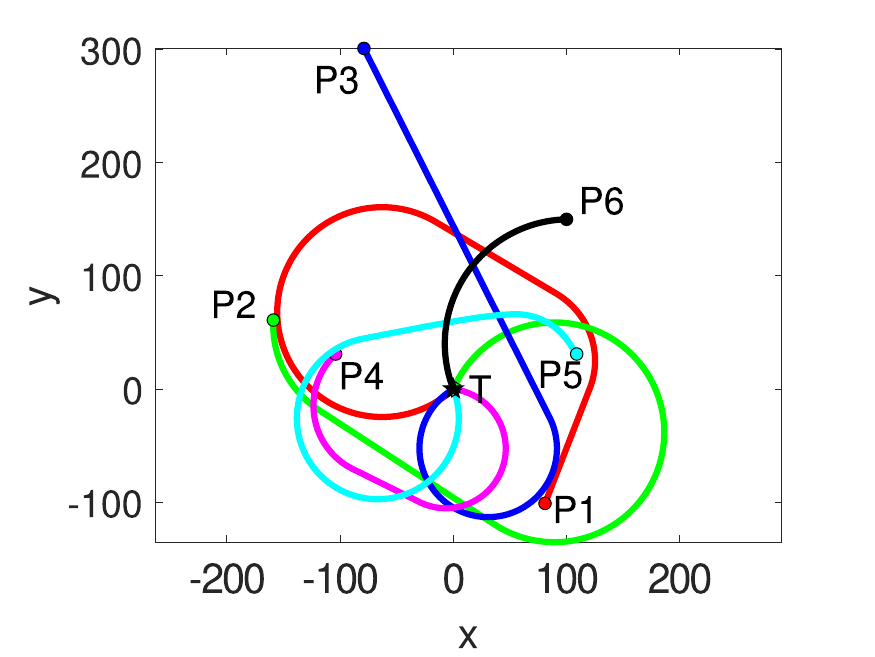}
        \caption{Case $3$}
        \label{fig:Trajectories_Case4}
     \end{subfigure}
        \caption{The trajectories of pursuers are shown, with $P1,P2,P3,P4,P5$ denoting the starting points of the respective pursuers. In subplot (d), a new pursuer $6$ is added to the graph after $1$ second.}
\end{figure*}
\begin{figure*}[ht]
     \centering
     \begin{subfigure}[b]{0.23\textwidth}
        \centering
        \includegraphics[width=\linewidth,center]{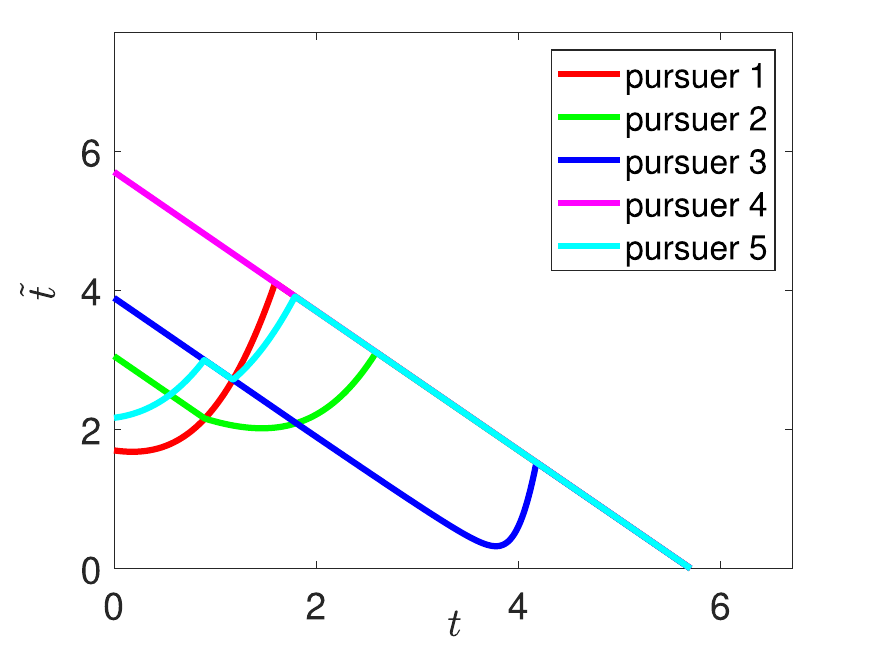}
        \caption{Case $1$}
        \label{fig:Case1_tilda}
     \end{subfigure}
     \hfill
     \begin{subfigure}[b]{0.23\textwidth}
        \centering    
        \includegraphics[width=\linewidth,center]{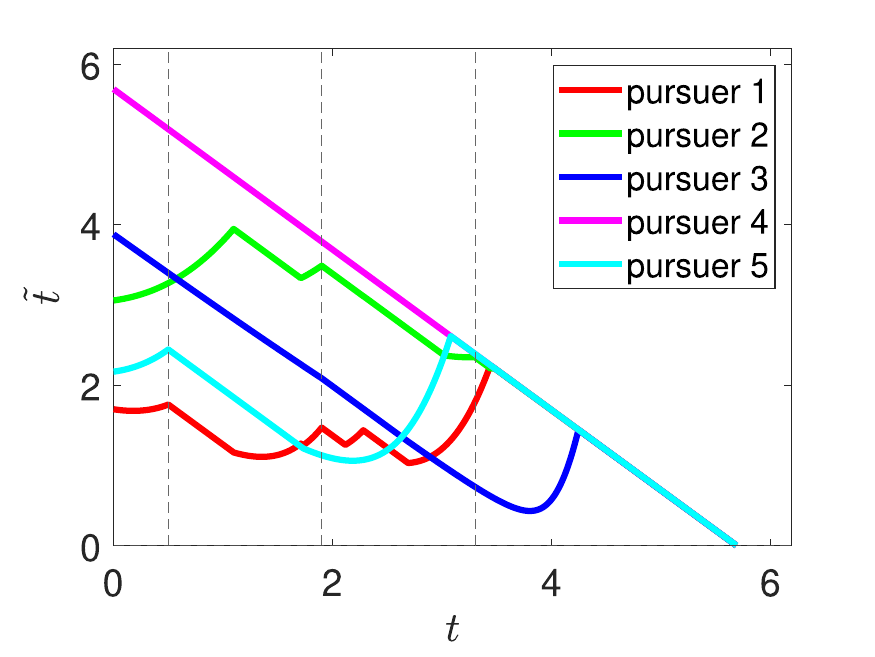}
        \caption{Case $2$ (without sinks)}
        \label{fig:tilda_Case2}
     \end{subfigure}
     \hfill
     \begin{subfigure}[b]{0.23\textwidth}
        \centering 
        \includegraphics[width=\linewidth,center]{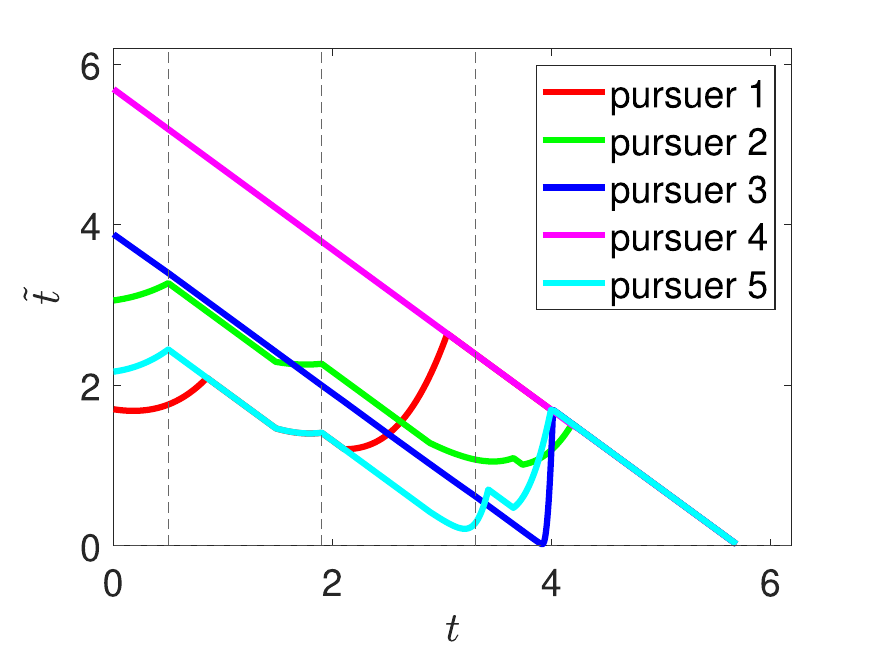}
        \caption{Case $2$ (with sinks)}
        \label{fig:tilda_Case3}
     \end{subfigure}
     \hfill
        \begin{subfigure}[b]{0.23\textwidth}
        \centering 
        \includegraphics[width=\linewidth,center]{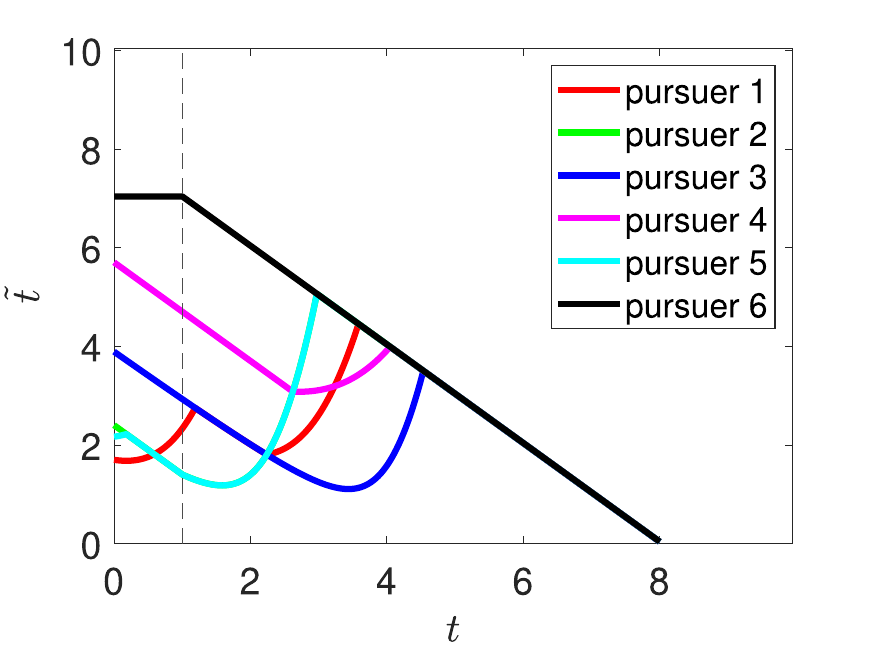}
        \caption{Case $3$}
        \label{fig:tilda_Case4}
     \end{subfigure}
        \caption{The plots of $\tilde{t}$ vs $t$ are shown. In subplots (b) and (c), vertical dotted lines indicate the time instants when the graph changes from $\mathcal{G}^c(t)$ to $\mathcal{G}^d(t)$. In subplot (d), a vertical dashed line marks the time instant when a new node (pursuer $6$) is added to the graph.}
\end{figure*}
and switching graphs, the following discussion entails how we can control the time of impact.
\subsection{Feasible desired times of impact}
The time of interception is always lower bounded by a finite value due to practical constraints like speed limitations and actuator saturation. For static graphs, we know from Theorems \ref{Theorem_consensus_static} and \ref{theorem:consensus_dynamic} that the interception time is governed by the leader. So, by controlling the its estimated time of interception, any interception time greater than the lower bound can be achieved using the proposed guidance law. For example, in scenarios where the launch angle $\gamma_l$ can be changed, eqn. \eqref{eq:t_tilde_i} can be used to impose a desired interception time. The same approach also works for switching graphs given that the leader does not change throughout the pursuit. 

In the next section, we present numerical simulations to illustrate all the theoretical results discussed in this paper. 
\section{Simulation results}
\label{Section5}
Consider a stationary target located at the origin. There are five pursuers located initially at $(81m,-101m)$, $(-159m,61m)$, $(-79m,301m)$, $(-104m,31m)$, $(109m,31m)$ with lead angles of $60^\circ$ ,$72^\circ$, $-12^\circ$ ,$120^\circ$ ,$72^\circ$ and speeds of $92m/s$, $73.6m/s$, $80.5m/s$, $46m/s$, $69m/s$, respectively. We present simulations for both static and switching graph topologies. In all figures, the yellow node represents the leader. Unlike the switching graphs shown in Fig. \ref{fig:gc_gd_figures}, we display only the first graph in $\mathcal{G}^c_{m-1}(t)$ and $\mathcal{G}^d_m(t)$ at each $m^{th}$ switch, as these graphs control the $\tilde{t}$ dynamics. The following case presents simulation results for simultaneous interception in static graphs.
%

%
\subsection*{Case $1$: Simultaneous interception in static graphs}
\begin{figure}[H]
     \centering
     \begin{subfigure}[b]{0.35\linewidth}
     \centering
     \includegraphics[width=0.65\linewidth]{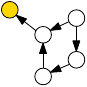} 
     \caption{A static graph}
     \label{fig:Figure_static_case_1}
     \end{subfigure}
     \hfill
     \begin{subfigure}[b]{0.55\linewidth}
    \centering 
    \includegraphics[width=0.82\linewidth]{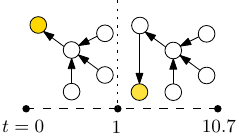}
    \caption{A new node is added at $t=1$}
    \label{fig:node_addition}
     \end{subfigure}
     \caption{Underlying graph topologies}
\end{figure}

%
Consider the pursuers connected over the static graph as shown in Fig. \ref{fig:Figure_static_case_1}. Under the proposed guidance law \eqref{eq:control_law}, the pursuers achieve simultaneous target interception, as stated
in Theorem \ref{Theorem_consensus_static}. This follows from the trajectories in Fig. \ref{fig:Case1}. Fig. \ref{fig:Case1_tilda} shows that the consensus in $\tilde{t}$s occurs in finite time.
%
Upon confirming simultaneous interception in static graphs, the following case establishes its verification in switching graphs, building on Theorems \ref{theorem:consensus_dynamic} and \ref{Theorem:consensus_const_sink}.
\subsection*{Case $2$: Simultaneous interception in switching graphs}
\label{subsection:case2}
\begin{figure}[ht]
    \centering 
    \includegraphics[scale=1,center]{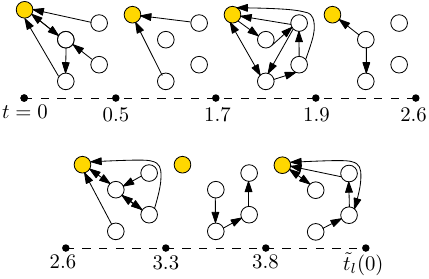}
    \caption{A series of graphs switching from $\mathcal{G}^c(t)$ to $\mathcal{G}^d(t)$ modes at times $t_m=\{0.5,1.9,3.3\}$, with $\delta_m=\{1.2,0.7,0.5\}$.}
    \label{fig:Dynamic_graphs_Case2}
\end{figure}

Consider the pursuers connected over the switching graphs shown in Fig. \ref{fig:Dynamic_graphs_Case2}. As per Theorem \ref{theorem:consensus_dynamic}, Fig \ref{fig:tilda_Case2} demonstrates that the consensus in $\tilde{t}$s is reached in finite time, resulting in simultaneous interception of the target. The vertical dotted lines at $t=\{0.5,1.9,3.3\}$ denote the instants when the graph switches from $\mathcal{G}^c(t)$ to $\mathcal{G}^d(t)$. At each switch, $\delta_m$ is set to $0.7\min_{i\in\mathcal{V}}~\tilde{t}_i(t_m)$, yielding $\delta_m=\{1.2,0.7,0.5\}$, as shown in Fig. \ref{fig:Dynamic_graphs_Case2}. The pursuers' trajectories are illustrated in Fig \ref{fig:trajectory_case_2}.
\begin{figure}[ht]
    \centering 
    \includegraphics[scale=1,center]{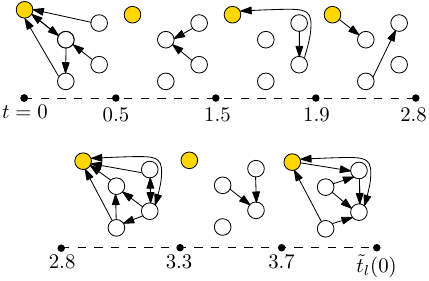}
    \caption{A series of graphs switching from $\mathcal{G}^c(t)$ to $\mathcal{G}^d(t)$ modes at $t_m=\{0.5,1.9,3.3\}$, with $\delta_m=\{1.7,0.9,0.4\}$.}
    \label{fig:Dynamic_Graphs_Case3}
\end{figure}
Now consider another scenario of switching graphs in which the sinks and singletons remain unchanged within each switch $m$ in $\mathcal{G}^d_m(t)$, as shown in Fig. \ref{fig:Dynamic_Graphs_Case3}.  Theorem \ref{Theorem:consensus_const_sink} and Fig \ref{fig:tilda_Case2} confirms the simultaneous interception of the target. The following key observations in this scenario include:
\begin{itemize}
    \item The communication graph is identical to the previous scenario for $t\in[0,0.5)$, resulting in the same dynamics. 
    \item At $t=0.5$, the disconnected graph $\mathcal{G}_1^d(t)$ begins, with $\delta_m$ calculated as $1.7~s$ (using a similar 0.7 factor) according to Theorem \ref{Theorem:consensus_const_sink}, which is higher than in the previous case. 
    \item The graph changes at $t=1.5~s$ without altering the sink nodes, leaving the overall dynamics unaffected.
\end{itemize}
 Next at $t=1.9~s$, a new disconnected $\mathcal{G}_2^d(t)$ appears, and $\delta_m$ is recalculated to be $0.9~s$. These switches continue for the necessary duration of $\delta_m$ until consensus is achieved.
The pursuers' trajectories are shown in Fig \ref{fig:Trajectories_Case3}.

%
The results presented in this case clearly explain how simultaneous interception occurs as the graph switches over time. The following case further illustrates this concept by adding nodes in a static graph as per Corollary \ref{cor:addition/removal}.
\subsection*{Case $3$: Simultaneous interception in switching graphs with node additions}
%

Consider the case where a node initially positioned at $(100m,150m)$ with lead angle of $55\text{\textdegree}$ and speed of $30m/s$ is added to the static graph at $t=1$, as depicted in Fig. \ref{fig:node_addition}. The addition follows Corollary \ref{cor:addition/removal}. The corresponding trajectories are displayed in Fig. \ref{fig:Trajectories_Case4}. The pursuers reach consensus in $\tilde{t}$s in finite time, as shown in Fig. \ref{fig:tilda_Case4}.
%

\section{Conclusion}
\label{Section6}
In this paper, we consider a system of $n$ heterogeneous pursuers modelled as unicycles having constant speeds. A max-consensus based distributed guidance law is proposed for the simultaneous interception of a stationary target under Assumption \ref{assump:not_0_and_pi}. When the latter does not hold, Remark \ref{rem:1} can be used to circumvent the issue. In the given framework, we designate a node with the largest $\tilde{t}$ as the leader. For static graphs, we prove that simultaneous interception of a stationary target is guaranteed if the leader remains globally reachable at all times. When the graphs are switching in nature, we categorise the switched graphs as either $\mathcal{G}^c(t)$ or $\mathcal{G}^d(t)$ pertaining to the global reachability of the leader or the lack of it, respectively. Thereafter, we determine the necessary conditions on the duration of the existence of $\mathcal{G}^d(t)$ (as discussed in eqn. \eqref{eq:dyn_graph_discontinuity}), which ensure simultaneous interception even if the leader loses its global reachability. Through suitable adjustments of the initial conditions, we also outline the necessary conditions required to impose a desired time of interception.

In Sec. \ref{Section5}, we present some simulation results to illustrate and validate the results discussed in the paper. Through Case $3$, we emphasise the fact that the leader can change during the pursuit, while the finite time simultaneous interception does not get affected. Future works in this direction may extend this approach to limit the lateral acceleration required throughout the engagement and to address the interception of a manoeuvring target.

\addtolength{\textheight}{-12cm}   


\bibliographystyle{ieeetran}        
\bibliography{main}

\end{document}